# Coherent direction-dependent valley transport in Dirac/Weyl semimetals

Can Yesilyurt[a)]

*Nanoelectronics Research Center, Istanbul, Turkey*

We investigate the coherent transmission of collimated wave packets through an inversion-asymmetric electrostatic barrier in a Dirac or Weyl channel. For a barrier consisting of a vertical and an oblique interface, we show that forward- and backward-moving wave packets experience different Fermi-surface mismatch sequences, causing direction-dependent transport properties in channels with both isotropic and anisotropic energy dispersion. Furthermore, the simulation results reveal that the same barrier structure leads to valley-resolved direction-dependent transmission in tilted anisotropic systems. The proposed device is fully electrostatic, gate-tunable, and, in principle, compatible with a wide range of materials hosting Dirac/Weyl fermions.

## I. INTRODUCTION

Control over the direction of electronic transport is of considerable fundamental and technological importance. Conventional approaches rely on breaking inversion or time-reversal symmetry in the host material, as demonstrated in polar crystals,[1] non-centrosymmetric superconductors,[2,3] or systems exhibiting magnetic order,[4–6] or they require the application of external fields. In two-dimensional (2D) Dirac and Weyl materials,[7–9] such symmetry breaking is frequently absent or undesirable. For instance, pristine graphene and gapless type-I tilted Dirac cones are inversion-symmetric, and most established mechanisms for directional transport require additional magnetic barriers or strain.[10,11] Achieving a carrier-selective directional asymmetry, in which the transmission of charge, spin, or valley currents across a single device depends on the direction of injection, remains one of the primary goals in 2D Dirac materials. Recent developments include composite metal-on-topological-insulator stacks, where one chiral edge state is hybridized away, resulting in nearly perfect directed charge and spin transport at zero bias,[12] as well as Floquet engineering with circularly polarized light to realize electronic isolators and zero bias spin-polarized photocurrents in laser-illuminated graphene and bilayer graphene.[13,14]

In multivalley band structures, charge carriers possess a binary valley quantum number, which has become a valuable resource for both classical and quantum information processing.[15–19] Valley-dependent electron optics has been thoroughly explored at single barrier interfaces in tilted Dirac and Weyl systems, where the tilt produces differential refraction of $K$ and $K'$ wave-packets, enabling valley filtering and beam splitting.[20–26] Tilted type-I Dirac cones are predicted in 8-$Pmmn$ borophene ($\zeta \approx 0.46$).[27,28] Experimental evidence for tilted cones has been reported in the layered organic conductor $\alpha$-(BEDT-TTF)$_2$I$_3$ under pressure[29] and in the Weyl semimetal NbP.[30] Optical conductivity measurements of the type-II Weyl candidate TaIrTe$_4$ yield an effective type-I tilt.[31] Dirac-cone tilt can also be engineered artificially in graphene superlattices.[32] The tilt $\zeta_y = 0.35$ used throughout is representative of this class (see Sec. III).

Direction-dependent transmission of ballistic carriers through inversion-asymmetric geometries is established for conventional two-dimensional electron gases[33] and has been studied with Dirac wave packets in geometrically asymmetric graphene structures.[34] Time-dependent propagation reproduces Klein tunneling and caustics at graphene junctions.[35,36]

This study demonstrates that an inversion-asymmetric electrostatic barrier shape alone can induce coherent direction-dependent transmission in a Dirac or Weyl channel with isotropic dispersion. Introducing a Dirac-cone tilt produces valley-resolved directional transmission, with dichroism that reverses sign at the Dirac point. The proposed device is fully electrostatic, gate-tunable, and, in principle, compatible with coherent ballistic Dirac and Weyl channels that support gate-defined barriers and finite-aperture injection. The observable is the transmission of prepared coherent beams for the two directions of injection, including a normally incident Gaussian packet, as well as a coherent angular source spanning approximately 30°. Normal-incidence transmission is a standard figure of merit for gate-defined electron-optical devices,[37] since normal incidence dominates the current driven between source and drain[38] and gate-defined barriers transmit predominantly near-normal carriers.[39] The beam width determines its angular acceptance, with broader beams being more collimated and enhancing directional contrast, as quantified in the supplementary material. Reciprocity relates transmission amplitudes upon exchange of reciprocal channels. The total transmission of a selected incident beam can therefore depend on direction, while complete channel sums with matched weights obey the reciprocity relation.[40,41] Section II F analyzes the channel structure of the transmission.

[a)]Author to whom correspondence should be addressed: cyesil@nanorescenter.com; can--yesilyurt@hotmail.com

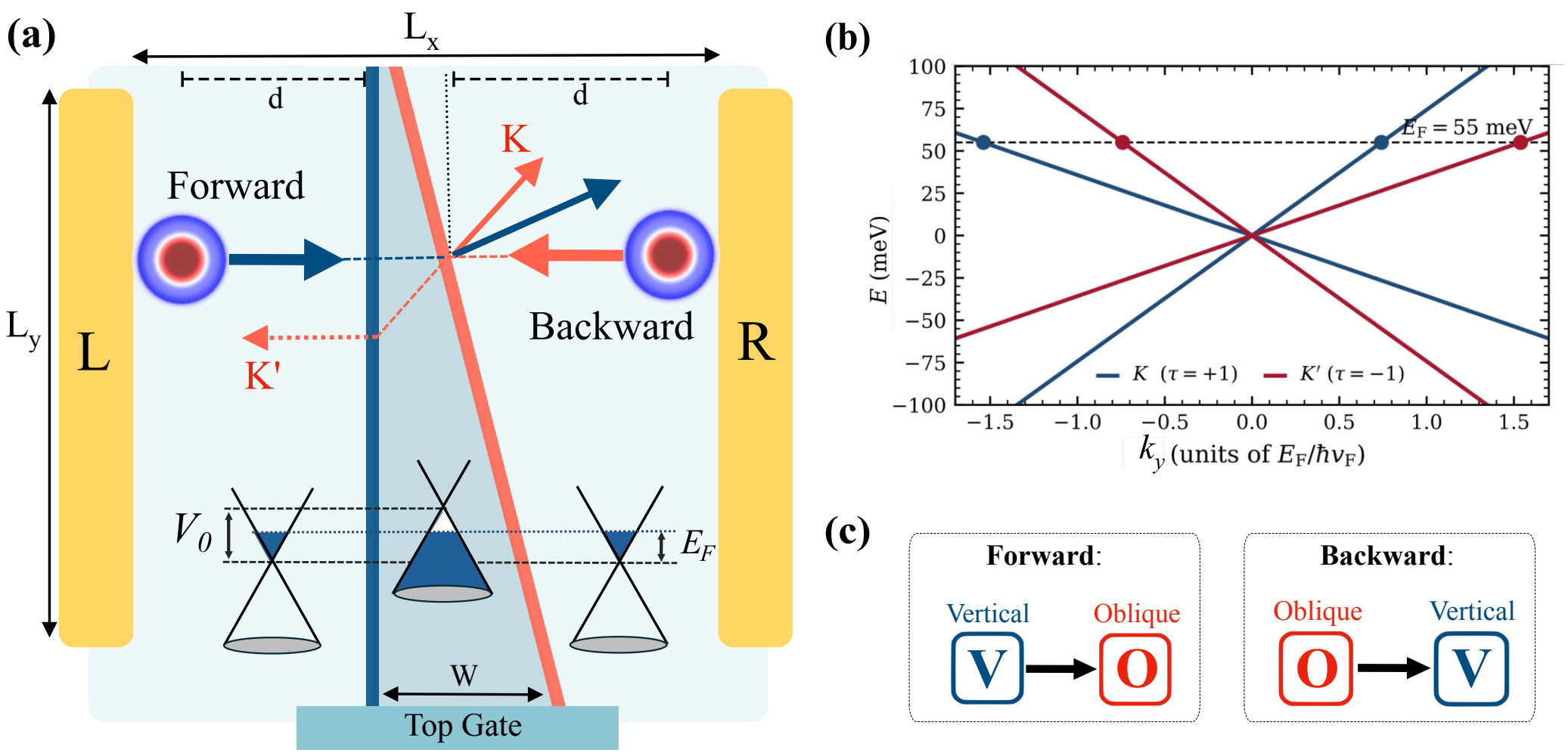

FIG. 1. **Device concept and direction-dependent interface composition. (a)** A top gate defines an inversion-asymmetric right-angle barrier, with one vertical ($V$) and one oblique ($O$) face, on a two-dimensional Dirac/Weyl channel ($L_x \approx L_y \approx 3.3$ µm); the leads L and R sit at equal distances $d = 698\, nm$ from the barrier. The barrier has base width $W = 500\, nm$ and width $W/2 = 250\, nm$ at $L_y/2$, the mid-channel axis along which Gaussian wave-packets of width $\sigma = 132\, nm$ are injected. The Dirac cones depict the energy dispersion and carrier filling inside and outside the barrier region. **(b)** Two-valley low-energy band structure with valley-opposite Dirac-cone tilt $\zeta_y = w_y/v_\mathrm{F} = 0.35$; dots mark the Fermi-energy intercepts at $E_\mathrm{F} = 55\, meV$. **(c)** Interface-composition diagram for the two injection directions.

## II. RESULTS

### A. Model and device

We consider a two-dimensional Dirac/Weyl channel of finite width described, near a pair of valleys $\tau = \pm 1$, by the tilted Dirac/Weyl Hamiltonian

$$\mathrm{H}_\tau = \tau\hbar\, \mathrm{w}\cdot\mathrm{k}\,\mathbb{I} + \hbar v_\mathrm{F}(\sigma_\mathrm{x}\mathrm{k}_\mathrm{x} + \tau\sigma_\mathrm{y}\mathrm{k}_\mathrm{y}) + \mathrm{V}(\mathrm{x},\mathrm{y})\,\mathbb{I} + \mathrm{M}(\mathrm{y})\,\sigma_\mathrm{z}, \quad (1)$$

where $\sigma_i$ are Pauli matrices in the sublattice pseudospin basis, $\mathbb{I}$ is the $2\times 2$ identity, $\boldsymbol{w} = (0, w_y)$ is the transverse Dirac-cone tilt, $V(x,y)$ is the gate-defined electrostatic potential, and the mass term $M(y)\,\sigma_z$ implements an infinite-mass confinement on the channel side walls.[42,43] Figure 1 shows the device model where two leads contact the channel at the same distance $d$ from the barrier on either side, and a top gate defines an electrostatic barrier $V_0$ shaped as a right-angle barrier with a vertical face and an oblique face [Fig. 1a]. The tilt enters as a shift proportional to the identity, modifying the propagation direction of each valley but leaving the cone structure, and hence the energy degeneracy at the Dirac point, intact; time-reversal symmetry imposes the sign reversal $\boldsymbol{w} \to -\boldsymbol{w}$ between the two valleys, $K$ and $K'$,[8,9] which shifts their constant-energy contours oppositely along $k_y$ [Fig. 1b]. The mechanism originates in the way the two qualitatively distinct interface types act on this valley-resolved carrier momentum in a direction-dependent manner [Fig. 1c]. The ratio $\zeta_y = w_y/v_\mathrm{F}$ is the type-I tilt parameter; throughout, we take $v_\mathrm{F} = 10^6\ \mathrm{m\,s^{-1}}$, unless otherwise stated. The scattering problem depends on $E_\mathrm{F}$ only through $V_0/E_\mathrm{F}$, the barrier width in Fermi wavelengths and the collimation $\sigma k_\mathrm{F}$. Section II C presents the dependence on $V_0/E_\mathrm{F}$ over the full range, including negative $V_0$.

The electrostatic barrier, $V(x,y) = V_0$ inside a region bounded by two linear interfaces and zero outside, defines the device shape. We compare four barrier shapes in the same tilted channel: (i) a rectangular barrier with two vertical interfaces (V–V); (ii) an isosceles triangle with two oblique interfaces related by reflection in the transport coordinate (O–O, mirror-symmetric); (iii) a right-angle barrier with one vertical and one oblique interface (V–O); and (iv) a parallelogram barrier with two parallel oblique interfaces (O–O, parallel), whose faces make the same angle with the channel axis as the oblique face of the right-angle barrier. The right-angle barrier is also studied with isotropic dispersion ($w_y = 0$) to isolate the role of geometry.

In a uniform region with constant potential $V$, the eigenenergy and eigenspinor of valley $\tau$ in band $s = \pm 1$ are

$$\varepsilon_{s,\tau}(\boldsymbol{k}) = \tau\hbar\,\boldsymbol{w}\cdot\boldsymbol{k} + V + s\,\hbar v_\mathrm{F}\,|\boldsymbol{k}|,$$
$$\chi_{s,\tau}(\boldsymbol{k}) = \frac{1}{\sqrt{2}}\begin{pmatrix}1\\ s\,e^{i\tau\phi_{\boldsymbol{k}}}\end{pmatrix}, \quad \phi_{\boldsymbol{k}} = \arg(k_x + ik_y), \quad (2)$$

with $s = +1$ for the electron-like states in the source region and $s = \mathrm{sgn}(E_\mathrm{F} - V_0)$ inside the barrier, so that the hole-like ($V_0 > E_\mathrm{F}$, Klein) regime is correctly described by the lower-band spinor. Setting the electron-band eigenenergy of Eq. (2), $\varepsilon_{+,\tau}(\boldsymbol{k})$, equal to $E_\mathrm{F}$ in the source region ($V = 0$) determines the Fermi wave vector at incidence angle $\phi$:

$$k_{\mathrm{F},\tau}(\phi) \;=\; \frac{E_{\mathrm{F}}}{\hbar\left(v_{\mathrm{F}} + \tau\, w_y \sin\phi\right)}, \tag{3}$$

so the source-side Fermi surface of $K$ is shifted toward $-k_y$, and that of $K'$ is shifted toward $+k_y$. Inside a barrier of height $V_0$ the longitudinal wavevector is

$$q_{x,\tau}(V_0, k_y) \;=\; \frac{1}{\hbar v_{\mathrm{F}}}\sqrt{(E_{\mathrm{F}} - V_0 - \tau\hbar w_y k_y)^2 - (\hbar v_{\mathrm{F}} k_y)^2}, \tag{4}$$

with the usual analytic continuation $q_{x,\tau} \to i|q_{x,\tau}|$ in the evanescent regime, where the argument of the square root is negative. For the central $k_y = 0$ component, the condition $V_0 = E_{\mathrm{F}}$ places the barrier interior at the Dirac point ($q_{x,\tau} \to 0$); the finite angular and energy spread of the injected beam broadens this condition into a window around $E_{\mathrm{F}}$.

The transmission characteristics reported below are obtained by solving the time-dependent Dirac equation $i\hbar\, \partial_t \Psi = H_\tau \Psi$ for each valley $\tau$ by direct wave-packet propagation. For normal incidence, a normalized Gaussian spinor wave-packet of width $\sigma$ is launched from the source side with central momentum $\boldsymbol{k}_0$ aligned along $\hat{\boldsymbol{x}}$, and the probability absorbed at the drain opposite to the source defines the cumulative transmission $T_\tau$, while that absorbed on the source side defines the reflection $R_\tau$; the propagation scheme, boundary conditions, and absorber masks are described in the supplementary material in detail. The packet width $\sigma$ is the effective transverse width of the injected beam at the contact and can be set experimentally by the injector and collimator design.[44,45] It determines the angular acceptance $\Delta\theta = 1/(2\sigma k_{\mathrm{F}})$, equal to 2.6° for the parameters of Fig. 1, and the energy resolution $\hbar v_{\mathrm{F}}/(2\sigma) = 2.5$ meV (see supplementary material for the dependence of the transmission on $\sigma$). We refer to this as normal incidence in the sense of normal central *momentum*; in the presence of tilted energy dispersion along the transverse direction, the corresponding group velocity acquires a transverse component $v_y = \tau w_y$; thus, the central ray is itself valley-dependent. Forward and backward runs use the same barrier and the same wave-packet, interchanging the source position and the sign of $k_{x,0}$. Comparing forward and backward injection within the same valley characterizes the direction dependence of the beam transmission. For each injection direction, we report the normalized transmission $T_\tau/(T_\tau + R_\tau)$, and define the valley polarization $\eta = (T_K - T_{K'})/(T_K + T_{K'})$, evaluated separately.

The absorbing contacts span the full channel width at both ends, and the side walls are lossless. The contacts collect transmitted and reflected components irrespective of exit angle. Probability accounting includes the accumulated transmission, the reflection and the probability remaining in the channel, $P_{\mathrm{rem}} = 1 - T_\tau - R_\tau$. The same contacts, walls and detectors are used in all simulations, including the barrier-free reference ($T \approx 0.998,\ R \approx 5 \times 10^{-4}$) and the mirror-symmetric controls of Sec. II E, whose forward and backward transmissions show no noticeable direction dependence for the rectangular and isosceles barriers over the applied potential range.

### B. Interface refraction and ordering

At any straight interface between two regions of different electrostatic potential, the momentum component tangent to the interface is conserved,

$$\boldsymbol{k}^{(\mathrm{src})} \cdot \hat{\boldsymbol{t}} \;=\; \boldsymbol{k}^{(\mathrm{dst})} \cdot \hat{\boldsymbol{t}}, \tag{5}$$

the Dirac analog of Snell's law.[39,46] The scalar potential conserves the tangential momentum and does not directly act in pseudospin space; the reflection and transmission amplitudes are nonetheless fixed by spinor matching of the chiral states across the interface, which is what governs the angular selectivity and Klein tunneling discussed below. At a vertical interface ($\hat{\boldsymbol{n}} = \hat{\boldsymbol{x}}$), $\hat{\boldsymbol{t}} = \hat{\boldsymbol{y}}$ and Eq. (5) reduces to

$$k_y^{(\mathrm{src})} = k_y^{(\mathrm{dst})}. \tag{6}$$

The tilt-induced shift of the Fermi surface along $k_y$ alters $k_x$ inside the barrier through Eq. (4) but leaves the conservation law itself unchanged. For any potential that depends on $x$ only, the reflection $y \to -y$ maps $H_K$ onto $H_{K'}$, so that $T_{K'}(k_y) = T_K(-k_y)$. This relation permits valley polarization in $k$-space while preserving valley-degenerate transmission for an incident beam mirror symmetric about the transport axis and for the complete sum over modes. Thus, a rectangular barrier has valley-independent total transmission for such a mirror-symmetric preparation at any tilt.

An oblique interface behaves differently. For an interface tilted by $\alpha$ from the vertical, $\hat{\boldsymbol{n}} = (\cos\alpha, \sin\alpha)$ and $\hat{\boldsymbol{t}} = (-\sin\alpha, \cos\alpha)$, and Eq. (5) becomes

$$k_\parallel \;\equiv\; \boldsymbol{k} \cdot \hat{\boldsymbol{t}} \;=\; -k_x \sin\alpha + k_y \cos\alpha \;=\; \text{const.} \tag{7}$$

which couples $k_x$ and $k_y$; the conserved tangential component $k_\parallel$ depends on the source-side Fermi wavevector $k_{\mathrm{F},\tau}(\phi)$, itself valley-dependent through Eq. (3). Combining the two,

$$k_{\parallel,\tau}(\phi) \;=\; k_{\mathrm{F},\tau}(\phi)\,(-\cos\phi \sin\alpha + \sin\phi \cos\alpha), \tag{8}$$

so that for $w_y \neq 0$ the same incidence angle is mapped to a *different* destination-side wavevector for $K$ and $K'$. An oblique interface is therefore required for valley-polarized total transmission.

In an extended barrier, the carrier scatters at the first interface, propagates through the barrier interior, and scatters again at the second, so the transmission depends on the operator composition $\mathcal{S}_2 \circ \mathcal{S}_1$. Under forward injection ($V \to O$) the carrier meets the vertical face first, where $k_y$ is conserved valley-blindly, and then the oblique face, where the conserved $k_\parallel$ separates the valleys through Eq. (8); under backward injection ($O \to V$) the order is reversed, so the carrier meets the oblique face first and the valleys are already separated before it reaches the vertical face. The two compositions $\mathcal{S}_O \circ \mathcal{S}_V$ and $\mathcal{S}_V \circ \mathcal{S}_O$ project the source Fermi surface onto the barrier-interior wavevectors in a different order, so their evanescent-mode support and Fermi-surface-mismatch sensitivities differ. For an inversion-symmetric barrier shape, such as a rectangle with two parallel vertical faces or an isosceles triangle with two oblique faces related by $x \to -x$, a device symmetry maps the forward composition onto the backward one, and the forward and backward beam transmissions coincide. For the inversion-asymmetric V–O right-angle barrier, no such symmetry exists: $\mathcal{S}_O \circ \mathcal{S}_V \neq \mathcal{S}_V \circ \mathcal{S}_O$.

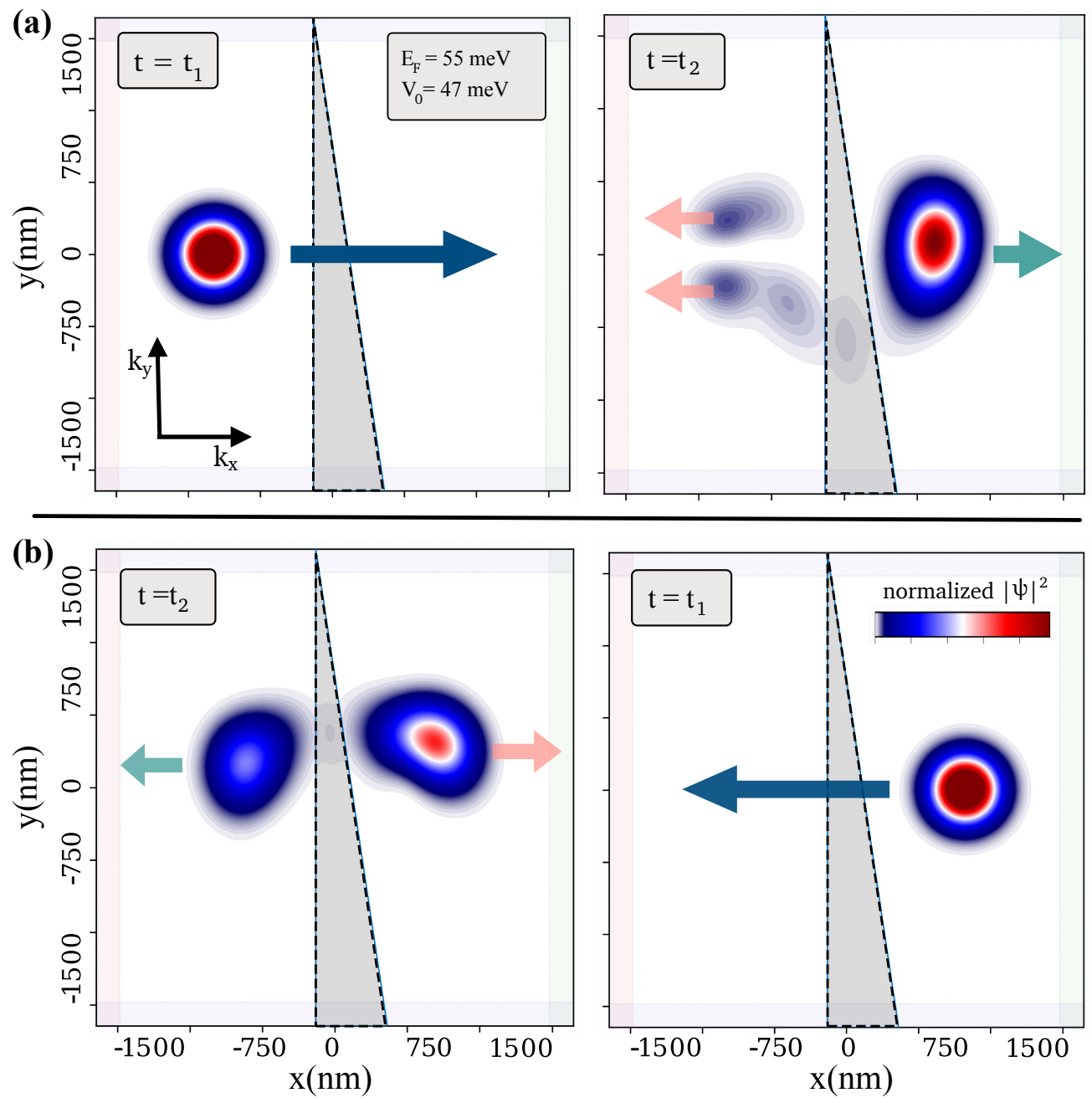

FIG. 2. **Direct visualization of coherent direction-dependent transmission.** Time-resolved snapshots of the wave-packet probability density $|\Psi|^2$ for forward and backward injection through the V–O barrier; the shaded wedge marks the barrier region, and $t_1$ and $t_2$ denote an early time, shortly after launch, and a late time, after scattering off the barrier. In each row, time evolves along the propagation direction. **(a)** Forward propagation ($V \to O$), a Gaussian packet launched from the left lead with central momentum along $+\hat{\boldsymbol{x}}$. **(b)** Backward propagation ($O \to V$), an identically prepared packet launched from the right lead with central momentum along $-\hat{\boldsymbol{x}}$. Other parameters as in Fig. 1.

Figure 2 visualizes the mechanism directly: the forward-injected packet emerges beyond the barrier as a compact, coherent transmitted lobe with only a weak reflected component, whereas the backward-injected packet is predominantly reflected, leaving a much smaller transmitted lobe; this direction-dependent asymmetry is quantified across the full $V_0$ range in the figures that follow.

### C. Direction-dependent transport with isotropic energy dispersion

The case $w_y = 0$ corresponds to a system in which the two valleys have isotropic energy dispersion and are degenerate in $k$-space, so the transport is purely charge-mode and any directional asymmetry of the charge transport originates in the broken interface-orientation symmetry of the barrier alone. The forward and backward coherent transmissions through the right-angle barrier on this channel with isotropic energy dispersion are shown in Fig. 3. The forward channel remains broadly transmissive over the full $V_0$ range, while the backward channel develops a deep transmission notch in the vicinity of $V_0 \sim E_{\mathrm{F}}$, producing a sizable directional asymmetry of the charge transport in the same window, quantified by the directional contrast $\Delta T = T^{\rightarrow} - T^{\leftarrow}$. The directional contrast peaks at $\Delta T \approx 0.69$ near $V_0 \approx 50$ meV, where $T^{\rightarrow} \approx 0.80$ and $T^{\leftarrow} \approx 0.11$, and the backward transmission reaches its minimum, $T^{\leftarrow} \approx 0.04$, near $V_0 \approx 56$ meV.

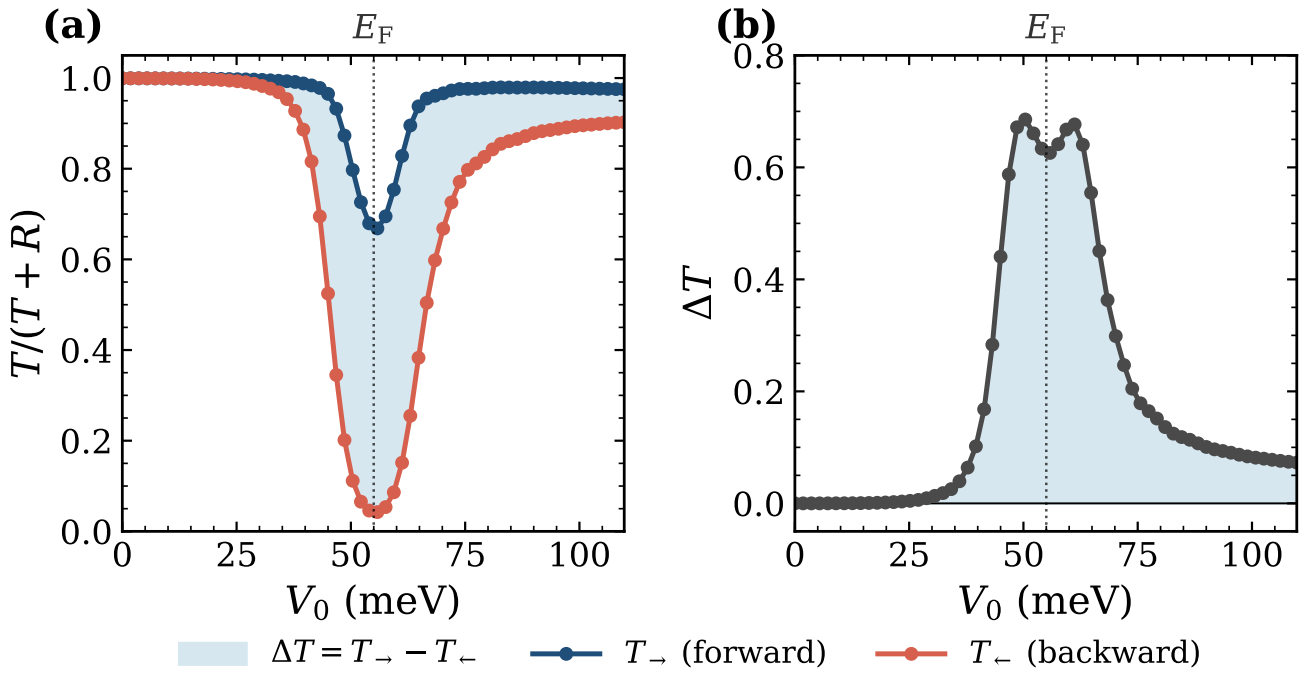

FIG. 3. **Direction-dependent charge-mode transmission in the absence of tilt.** Coherent transport through the right-angle barrier on a Dirac channel with *isotropic* energy dispersion ($w_y = 0$), plotted against the gate-defined barrier height $V_0$. **(a)** Normalized transmission $T/(T+R)$ for forward ($V \to O$) and backward ($O \to V$) injection. **(b)** Directional contrast of the charge transport, $\Delta T = T^{\rightarrow} - T^{\leftarrow}$. Device parameters as in Fig. 1, except $w_y = 0$ here.

In the absence of the barrier, the device is symmetric, so the directional contrast vanishes at $V_0 = 0$ by the symmetry of this reference geometry; it then grows as $V_0$ approaches the Fermi-surface-mismatch window, peaks near $V_0 \sim E_{\mathrm{F}}$, and decreases again once the barrier interior is pushed well away from the Dirac point. The peak occurs where the barrier interior sits near the Dirac point, and the available transmitted modes are most sensitive to the interface angle. The sign of $\Delta T$ follows from the barrier orientation: backward injection encounters the oblique face first, which couples a broader range of incident transverse momenta into evanescent or strongly deflected channels, thereby enhancing reflection.

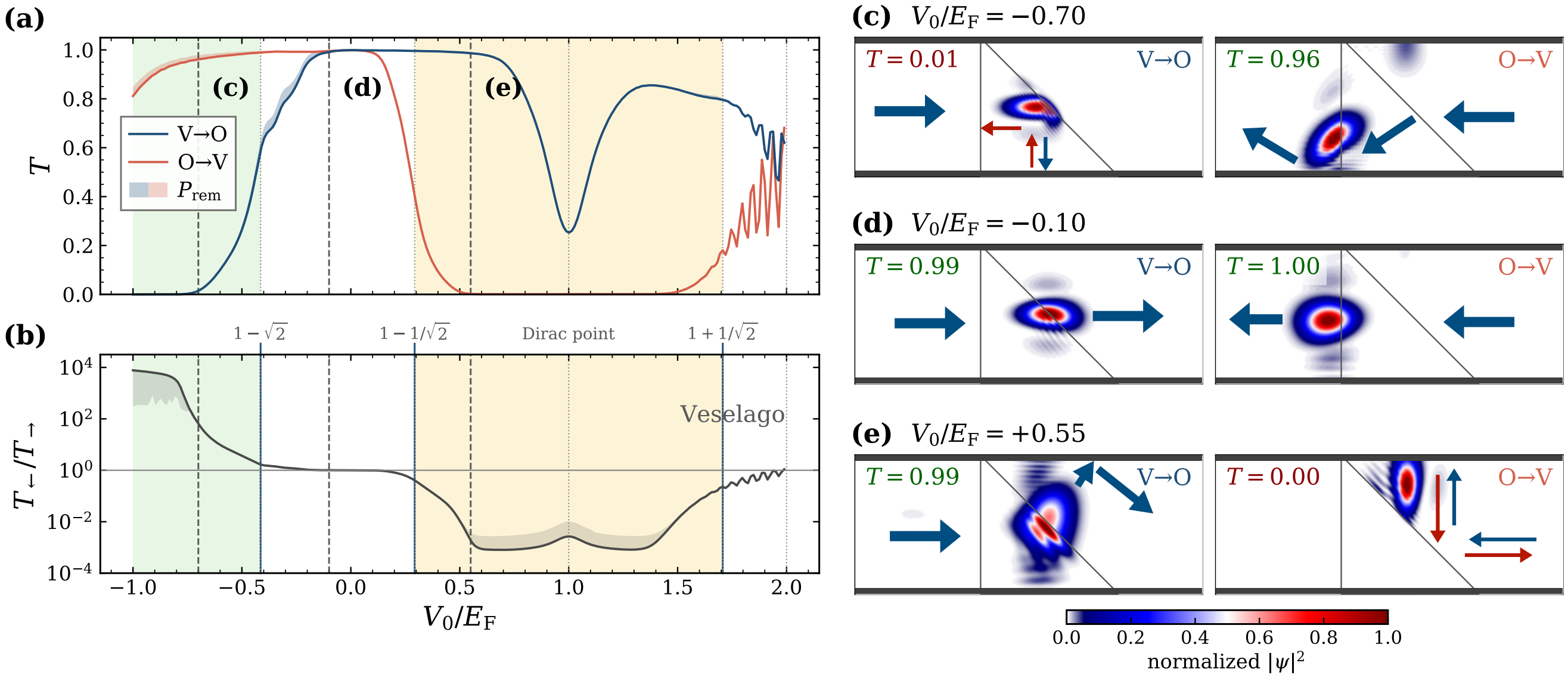

FIG. 4. **Operating regimes of the vertical–oblique barrier.** Isotropic dispersion, $E_F = 65$ meV and a 45° facet. **(a)** Contact-collected transmission for vertical-first (V→O, blue) and oblique-first (O→V, orange) injection. The source is the coherent superposition $\sum_n \psi(\phi_n)$ of eleven equal-amplitude packets with $\phi_n = 0°, \pm1.6°, \pm4.7°, \pm7.8°, \pm10.9°, \pm14.0°$, normalized and propagated as one wavefunction. Shaded bands show the uncollected probability $P_{\rm rem}$, drawn on top of the collected transmission $T$. **(b)** Ratio $T_\leftarrow/T_\rightarrow \equiv T_{O\to V}/T_{V\to O}$ on a logarithmic scale; the shaded band marks the bounds on the ratio set by the uncollected probability (supplementary Sec. SIV). Background shading distinguishes the central-ray blocking regimes; the critical-angle boundaries, Dirac point and Veselago point are marked. **(c–e)** Normalized probability density for $V_0/E_F = -0.70, -0.10$ and $+0.55$, one gate value in each regime of (a,b); the three regimes are tunable by $V_0$ and correspond to transport without direction dependence (d) and to strongly direction-dependent transport along one (c) or the other (e) direction.

The dependence on the Fermi energy can be expressed analytically. In the electron-optics description, each interface is a refractive step with relative index $n = |1 - V_0/E_F|$. For a normally incident beam and an oblique face at angle $\alpha$ to the vertical, the oblique-first direction is evanescent when $n < \sin\alpha$, while the vertical-first direction is blocked at the oblique exit when $n > 1/\sin\alpha$. The mechanism follows from the difference in interface orientations. A 45° facet gives the simple thresholds $1/\sqrt{2}$ and $\sqrt{2}$ and displays both blocking regimes within one gate range. Figure 4(a,b) shows the transmission of a coherent angular source spanning approximately 30° at 234 values of $V_0$ for $E_F = 65$ meV. It retains the three operating regimes predicted by the central-ray conditions, with transitions broadened by the source distribution. Oblique-first transmission dominates at sufficiently negative gate potentials, vertical-first transmission dominates around $V_0 = E_F$, and both directions transmit between and beyond these regimes. The gate thus controls both the magnitude and sign of the directional asymmetry. The shaded bands in Fig. 4(a) show $P_{\rm rem}$, the uncollected probability resonating in the channel. Figure 4(c–e) illustrates the corresponding wave-packet propagation. For the $\alpha = 8.6°$ barrier of Fig. 1, the oblique-first condition gives $47\ \text{meV} < V_0 < 63\ \text{meV}$, consistent with the backward notch in Fig. 3, broadened by the beam's angular spread. The blocking window is centered at $V_0 = E_F$ and has width $2E_F\sin\alpha$. The geometry, dataset and critical-angle derivation are given in supplementary Sec. SIV.

## D. Valley-resolved directional transport with anisotropic energy dispersion

Introducing a Dirac-cone tilt produces valley-resolved coherent directional transmission. For $\zeta_y = 0.35$, the four direction–valley channels follow distinct transmission curves within the same right-angle barrier geometry [Fig. 5a]. The forward channels remain nearly valley-independent, as the $K$ and $K'$ forward-transmissions closely follow each other with $|T_K^\rightarrow - T_{K'}^\rightarrow| \lesssim 0.10$ over the full $V_0$ range. In contrast, the backward channels display two pronounced transmission notches at separate barrier heights on either side of the Dirac point, thereby splitting the charge-mode notch observed in Fig. 3 into a valley-resolved doublet.

The per-valley directional contrast $\Delta T_\tau \equiv T_\tau^\rightarrow - T_\tau^\leftarrow$ [Fig. 5b] peaks for $K$ at a barrier height below $E_F$ and for $K'$ above $E_F$. These peaks fall on opposite sides of the Dirac point because the valley-dependent shift of the Fermi surface along $k_y$ drives the $K$ and $K'$ Fermi-surface-mismatch conditions at the oblique interface in opposite directions as $V_0$ varies. The per-direction valley contrast $T_K - T_{K'}$ [Fig. 5c] correspondingly remains near zero for forward injection but develops a sign-changing dipolar structure for backward injection, with a node at $V_0 = E_F$. The transmitted-current valley polarization [Fig. 5d] mirrors this, the backward direction switching from strong $K'$ polarization below $E_F$ to strong $K$ polarization above it. For the parameters of Fig. 5, the per-valley directional contrast reaches $\Delta T_K \sim 0.77$ and $\Delta T_{K'} \sim 0.79$ at their respective peak barrier heights, and the backward valley polarization $|\eta^{\leftarrow}|$ exceeds 0.6 on both sides of the Dirac point. The exact magnitudes depend on geometry and packet collimation, but the dichroic structure across $E_{\mathrm{F}}$ is robust.

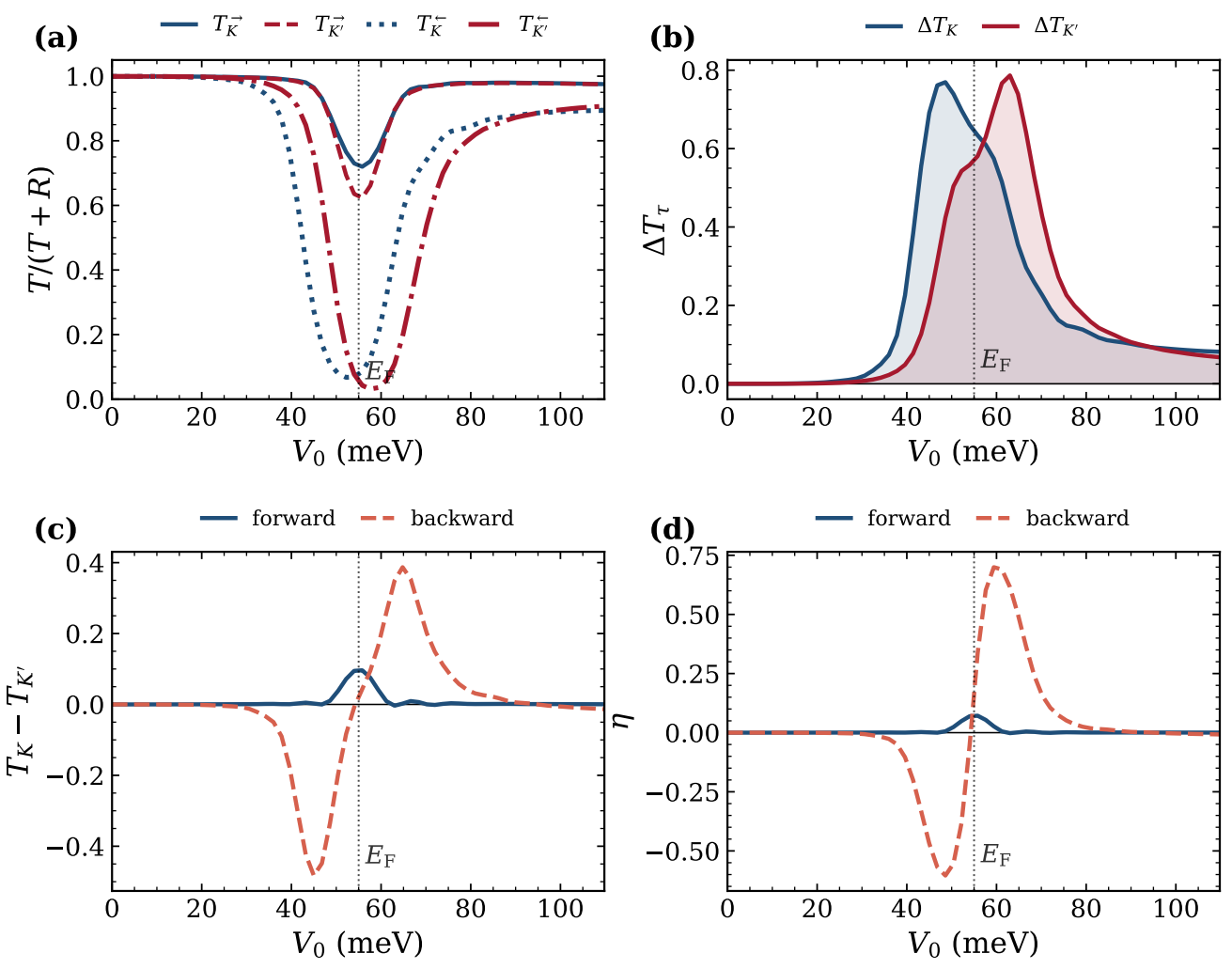

FIG. 5. **Valley-resolved directional transmission in the tilted right-angle barrier. (a)** Coherent transmission of the four direction×valley channels vs. barrier height $V_0$. **(b)** Per-valley directional contrast $\Delta T_\tau = T_\tau^{\rightarrow} - T_\tau^{\leftarrow}$ for the $K$ and $K'$ valleys. **(c)** Per-direction valley contrast $T_K - T_{K'}$ for forward and backward injection. **(d)** Transmitted-current valley polarization $\eta$ for forward and backward injection. Parameters as in Fig. 1.

The single control parameter $V_0$, set by a top gate, thus selects between a dominant-$K$ backward-blocking regime ($V_0 < E_{\mathrm{F}}$), a dominant-$K'$ regime ($V_0 > E_{\mathrm{F}}$), and an intermediate regime in which both valleys share the directional asymmetry: the device acts as a gate-switchable direction-dependent valley filter, with the valley set by the sign of the gate detuning $V_0 - E_{\mathrm{F}}$.

### E. Symmetry analysis of the shaped barrier

The geometric mechanism relates the directional transmission to the sequence of distinct interface orientations. We compare four barriers on the same tilted channel (Fig. 6, all with $\zeta_y = 0.35$): (a) a rectangular V–V barrier, (b) an isosceles O–O barrier with mirror-symmetric faces, (c) a right-angle V–O barrier, and (d) a parallelogram O–O barrier with parallel oblique faces, representing the barrier-rotation geometry.[47]

The rectangular V–V barrier [Fig. 6a] has valley-degenerate transmission ($|T_K - T_{K'}| \lesssim 10^{-5}$) and forward and backward curves that agree within 0.65% over the applied potential range. The isosceles O–O barrier [Fig. 6b] exhibits valley splitting up to $|T_K - T_{K'}| \sim 0.20$, while the two directions agree within 0.16% for each valley at every $V_0$. Reflection $x \to -x$ relates its forward and backward interface compositions. The right-angle V–O barrier [Fig. 6c] combines valley splitting with directional transmission because the two injection directions encounter inequivalent interface orientations in reverse order.

The parallelogram O–O barrier [Fig. 6d] shows the strongest valley splitting of the four geometries, with $|T_K - T_{K'}|$ up to 0.50. Its $C_2$ symmetry exchanges the valleys under direction reversal, giving $T_K^{\rightarrow} = T_{K'}^{\leftarrow}$ within 0.7% at every $V_0$. For equal injection into both valleys, the transmitted currents therefore sum to a direction-independent total, with the two directions agreeing within 0.11% over the applied potential range. Figure 7 shows that the right-angle barrier retains a charge-mode directional contrast of order unity. Barrier rotation controls valley selection, while inequivalent interface orientations encountered in a fixed order produce directional transmission of the total injected beam.

### F. Channel structure of the direction-dependent transmission

For the Hamiltonian in Eq. (1), time reversal $\Theta = \sigma_z \mathcal{K}$ maps $(K, \boldsymbol{k})$ to $(K', -\boldsymbol{k})$ and exchanges reciprocal input and output channels:

$$|t_{R\to L}^{K'}(-k_y' \to -k_y)|^2 = |t_{L\to R}^{K}(k_y \to k_y')|^2. \tag{9}$$

For a fixed incident channel $i$, the forward and backward totals are $T_i^F = \sum_j |t_{ji}|^2$ and $T_i^B = \sum_j |t_{ij}|^2$, respectively. These row and column sums can differ, while the complete sums over incoming modes are equal by unitarity. This distinction permits asymmetric transmission of selected incident modes in reciprocal systems.[40,41] Direction-dependent wave-packet scattering delays can likewise coexist with reciprocal transmission probabilities in ballistic two-terminal conductors.[48]

A flux-weighted transverse-momentum analysis at $V_0 = 50$ meV resolves the transmitted channels (supplementary Sec. SV and Fig. S2). The analysis uses $\sigma_x = 132$ nm and $\sigma_y = 460$ nm to improve transverse-momentum resolution. For $\zeta_y = 0.35$, normally incident $K$ injection from the left gives contact transmission 0.974, while normally incident $K'$ injection from the right gives 0.257. The forward output is deflected to $k_y \approx 0.0108\ \mathrm{nm}^{-1}$. Injecting the reversed output channel from the right in $K'$, along the reversed peak-bin direction and at the same central energy, $E = E_{\mathrm{F}} = 55$ meV, requires $k_y = -0.0109\ \mathrm{nm}^{-1}$ and $|k| = 0.0798\ \mathrm{nm}^{-1}$ and gives contact transmission 0.971. The corresponding integrated lobe weights, 0.89 and 0.90, agree at the percent level and are consistent with reciprocal channel conversion. These weights represent finite-packet channel groups rather than individual scattering-matrix elements. The reverse direction thus supports transmitting channels even when the normally incident preparation is strongly suppressed.

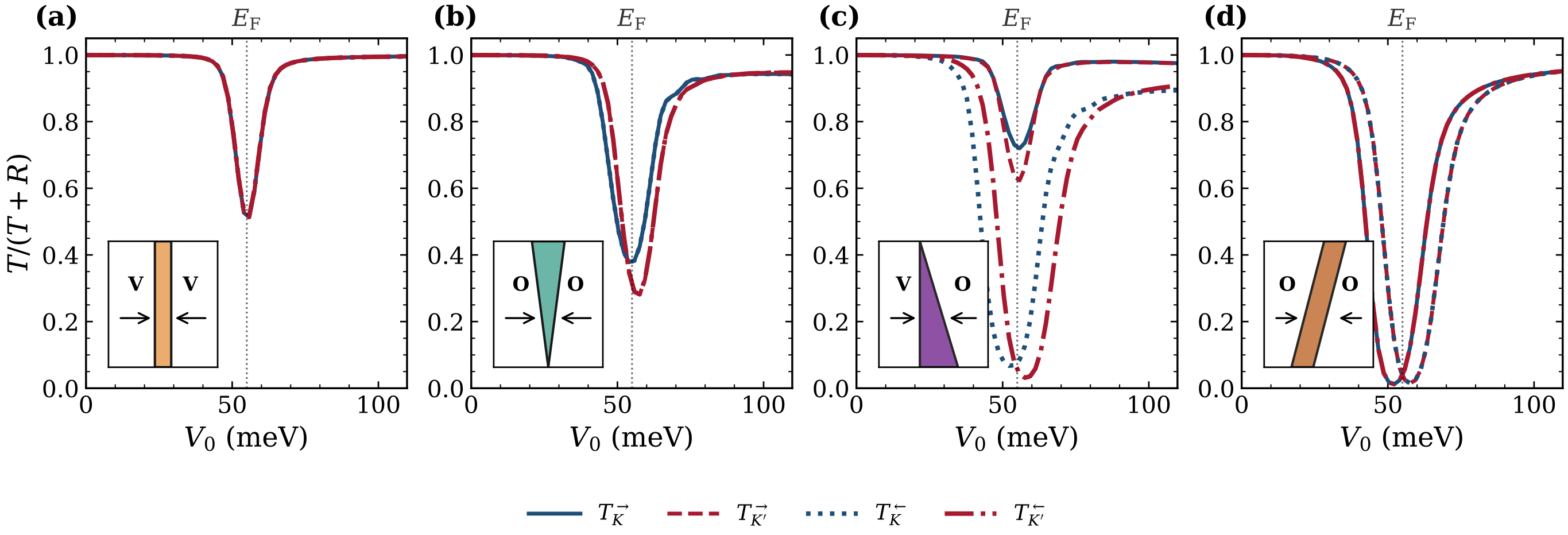

FIG. 6. **Symmetry analysis: four barrier shapes on the same tilted channel.** Each panel shows the coherent transmission $T/(T+R)$ vs. $V_0$ for the four direction–valley channels, with an inset of the barrier shape. **(a)** Rectangular V–V barrier. **(b)** Isosceles O–O barrier. **(c)** Right-angle V–O barrier. **(d)** Parallelogram barrier with two parallel oblique faces. Over the full $V_0$ range, the forward and backward transmissions of each valley agree within 0.65% in (a) and 0.16% in (b). In (d), the opposite-valley pairs, $T_K^{\rightarrow} = T_{K'}^{\leftarrow}$ and $T_{K'}^{\rightarrow} = T_K^{\leftarrow}$, agree within 0.7%. Relative differences use the smaller compared transmission. Other parameters as in Fig. 1.

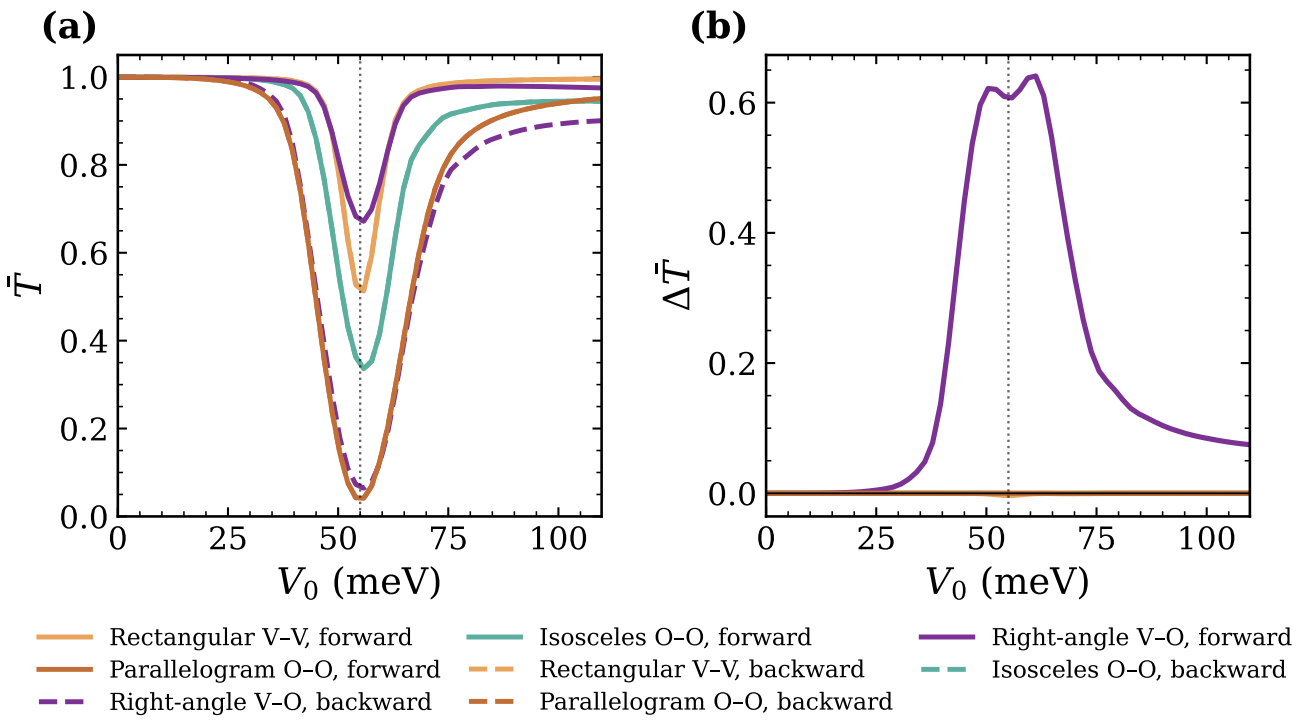

FIG. 7. **Charge-mode summary across the four tilted geometries.** For each barrier shape on the same tilted channel ($\zeta_y = 0.35$; other parameters as in Fig. 1), the valley-averaged transmission is $\bar{T} = (T_K + T_{K'})/2$. **(a)** Forward and backward charge-mode transmission. **(b)** Directional contrast $\Delta\bar{T} = \bar{T}^{\rightarrow} - \bar{T}^{\leftarrow}$. The right-angle V–O barrier produces a directional contrast. The forward and backward valley-averaged transmissions agree within 0.65%, 0.08% and 0.11% for the rectangular, isosceles and parallelogram barriers, respectively, relative to the smaller compared transmission.

## III. DISCUSSION AND CONCLUSIONS

The mechanism reported in this work provides a geometric route to coherent direction-dependent transmission in two-dimensional Dirac and Weyl materials. It operates with isotropic dispersion and follows from inequivalent interface orientations encountered in a direction-dependent order, yielding a non-commutative Fermi-surface-mismatch composition. The four-geometry comparison shows that only the right-angle triangular barrier combines valley splitting with directional transmission.

The proposed directional asymmetry of the charge transport can be achieved in any two-dimensional Dirac/Weyl channel, whereas its valley-resolved variant requires a finite in-plane Dirac-cone tilt. Over-tilted (type-II, $\zeta > 1$) materials reshape the Fermi-surface topology and fall outside the tilted-Dirac picture used here.

A range of platforms meets these requirements. The charge-mode device can be realized on pristine graphene, and the tilted platforms introduced in Sec. I, 8-*Pmmn* borophene,[27,28] $\alpha$-(BEDT-TTF)$_2$I$_3$ under pressure,[29,49,50] NbP[30] and graphene superlattices with engineered tilt,[32] support the valley-resolved variant.

Because the device is defined entirely by an electrostatic top gate, both the strength of the directional asymmetry and the valley selectivity are set by a single top-gate voltage, and the effect is within reach of existing gate-defined nanostructures on coherent ballistic Dirac or Weyl channels. Because the barrier width varies along $y$, the packet samples local widths between 210 and 290 nm across its $\pm 2\sigma$ extent. Displacing the packet toward the thin end increases transmission in the blocked direction, with directional contrast persisting at all positions examined [see the supplementary material (Fig. S3)].

*Note added in revision:* We became aware of a related work about asymmetric transmission through triangular electrostatic barriers in graphene.[51]

## SUPPLEMENTARY MATERIAL

See the supplementary material for wave-packet propagation and pseudospin preparation, boundary conditions and numerical parameters, injection collimation (Fig. S1), the energy-dependent regime map, channel-resolved transmission (Fig. S2), and injection-position and detector controls (Fig. S3).

## ACKNOWLEDGMENTS

This work was funded by the Nanoelectronics Research Center (Nano Electronics Research Mühendislik Araştırma Geliştirme Ltd. Şti.), Istanbul, Turkey.

## AUTHOR DECLARATIONS

### Conflict of Interest

The author has no conflicts to disclose.

### Author Contributions

C.Y. conceived the project, developed the simulation code, performed the simulations and the analysis, and wrote the manuscript.

## DATA AVAILABILITY

The numerical data and the simulation configuration files supporting the main findings of this study are openly available on Zenodo at https://doi.org/10.5281/zenodo.20612046 (ref. 52). The wave-packet simulations were performed with the open-source `dirac-wavepacket` package, version 1.0.1,[53] which is openly available on Zenodo at https://doi.org/10.5281/zenodo.19651770 with no access restrictions.

# Supplementary Material for:
# Coherent direction-dependent valley transport in Dirac/Weyl semimetals

Can Yesilyurt[a)]

*Nanoelectronics Research Center, Istanbul, Turkey*

## SI. WAVE-PACKET PROPAGATION

We use coherent time-dependent wave-packet propagation to resolve scattering through the sequence of interfaces. The two-component Dirac spinor evolves under the Hamiltonian of Eq. (1) in the main text, with the barrier potential, transverse confinement and absorbing contacts specified below.

For normal incidence, the initial state is a normalized Gaussian electron-like spinor wave-packet,

$$\Psi^{\text{in}}(\boldsymbol{r}) \;=\; \mathcal{N}\exp\left[-\frac{(x-x_0)^2}{4\sigma_x^2}-\frac{(y-y_0)^2}{4\sigma_y^2}\right]e^{i\boldsymbol{k}_0\cdot\boldsymbol{r}}\,\chi_{+,\tau}(\boldsymbol{k}_0), \tag{S1}$$

launched from the source side with central momentum $\boldsymbol{k}_0$ aligned along $\hat{\boldsymbol{x}}$ ($k_{y,0}=0$). The beam is centered at $y_0=0$, except in the injection-position scan of Sec. SVI. Preparing the electron-band eigenspinor at the central momentum aligns its pseudospin with $\boldsymbol{k}_0$. The angular spread gives a small hole-band admixture,[1,2] $P_{-}=\sum_{\boldsymbol{k}}|\tilde{\psi}(\boldsymbol{k})|^2\sin^2[(\phi_{\boldsymbol{k}}-\phi_{\boldsymbol{k}_0})/2]\approx 1/(16\sigma^2k_0^2)=5\times10^{-4}$ for $\sigma=132$ nm and $k_0=0.0836\ \text{nm}^{-1}$. This counter-propagating component is absorbed at the source contact, consistent with the barrier-free reflection $R\approx5\times10^{-4}$. The two-band propagator retains the associated zitterbewegung. The reported observables are the accumulated contact probabilities.

Probability absorbed at the drain mask opposite to the source accumulates as the cumulative transmission $T_\tau(t)$, and that absorbed by the source-side mask as the reflection $R_\tau(t)$. The propagation stops when the probability remaining in the channel, $P_{\text{rem}}=1-T_\tau-R_\tau$, falls below $10^{-3}$, or at the maximum propagation time of Table S1. The coefficients are evaluated at the end of propagation; the collection time for the calculations of Fig. 4 is specified in Sec. SIV. The probability balance is described in Sec. SII.

The two-component Dirac spinor is evolved with a second-order Strang split-operator Fourier scheme.[3–5] The Hamiltonian in Eq. (1) of the main text is split into a real-space part and a momentum-space part,

$$\begin{aligned}H_r &= V(x,y)\,\mathbb{I}+M(y)\,\sigma_z,\\ H_{k,\tau} &= \tau\hbar\,\boldsymbol{w}\cdot\boldsymbol{k}\,\mathbb{I}+\hbar v_{\text{F}}(\sigma_x k_x+\tau\sigma_y k_y),\end{aligned} \tag{S2}$$

and one time step is propagated as

$$\Psi(t+\Delta t)\approx e^{-iH_r\Delta t/2\hbar}\,\mathcal{F}^{-1}\,e^{-iH_{k,\tau}\Delta t/\hbar}\,\mathcal{F}\,e^{-iH_r\Delta t/2\hbar}\,\Psi(t), \tag{S3}$$

with local error $O(\Delta t^3)$.[4] The real-space half step is diagonal in the spinor basis,

$$\psi_\uparrow\to e^{-i(V+M)\Delta t/2\hbar}\,\psi_\uparrow,\qquad \psi_\downarrow\to e^{-i(V-M)\Delta t/2\hbar}\,\psi_\downarrow, \tag{S4}$$

while the momentum-space step is evaluated analytically as a $2\times2$ unitary at each $\boldsymbol{k}$,[5]

$$e^{-iH_{k,\tau}\Delta t/\hbar}=e^{-i\tau(\boldsymbol{w}\cdot\boldsymbol{k})\Delta t}\left[\cos(v_{\text{F}}k\,\Delta t)\,\mathbb{I}-\,i\sin(v_{\text{F}}k\,\Delta t)\,\frac{k_x\sigma_x+\tau k_y\sigma_y}{k}\right], \tag{S5}$$

with $k=\sqrt{k_x^2+k_y^2}$. The tilt term factors out as a scalar phase. The valley index enters both this phase and the chiral term multiplying $\sigma_y$.

[a)] Author to whom correspondence should be addressed: cyesil@nanorescenter.com; can--yesilyurt@hotmail.com

## SII. BOUNDARY CONDITIONS AND ABSORBING CONTACTS

The transverse channel walls are imposed through a mass profile

$$M(y) \;=\; M_{\mathrm{wall}} \max\left[\cos^2\left(\frac{\pi\, d_{\mathrm{bot}}}{2W_y}\right),\, \cos^2\left(\frac{\pi\, d_{\mathrm{top}}}{2W_y}\right)\right], \tag{S6}$$

inside edge strips of width $W_y$, with $d_{\mathrm{bot}} = y + L_y/2$ and $d_{\mathrm{top}} = L_y/2 - y$ and $M(y) = 0$ in the channel interior; the $M(y)\sigma_z$ term opens a local gap at the edges, confining the packet through a smooth mass boundary.[6,7] The wall mass is $M_{\mathrm{wall}} = 500$ meV, giving a local gap $2M_{\mathrm{wall}} = 1$ eV, and the ramp width is $W_y = 0.05L_y = 165$ nm.

The longitudinal contacts apply multiplicative amplitude masks $m(d) = 1 - S\cos^2[\pi d/(2W_x)]$ within strips of width $W_x$ at each $x$-boundary, with $m = 1$ outside the strips. Here $d$ is the distance from the respective boundary, $S = 0.08$ is the damping at that boundary and $W_x = 0.05L_x = 165$ nm. The operation $\Psi \to m(d)\Psi$ is applied once per time step after unitary propagation. It represents a negative imaginary potential $-i\Gamma(d)$, where $\Gamma(d) = -(\hbar/\Delta t)\ln m(d) \geq 0$. The probability removed per step, $(1 - m^2)|\Psi|^2$, is integrated over each strip and accumulated as transmission at the drain and reflection at the source. Thus, the absorbing contacts also serve as full-width detectors.

With lossless transverse walls, the probability balance is $T + R + P_{\mathrm{rem}} = 1$. At the end of a propagation, whether stopped by the $10^{-3}$ threshold or by the maximum propagation time, $P_{\mathrm{rem}}$ is dominated by long-lived resonances between the side walls and the barrier interfaces; for the calculations of Fig. 4 it is displayed as shaded bands (Sec. SIV). The barrier-free calculation gives $T \approx 0.998$ and $R \approx 5 \times 10^{-4}$, consistent with the small hole-band admixture in Sec. SI; reflection from the absorbers is at most of that order. Table S1 lists the parameters for the main geometry and its symmetry controls.

TABLE S1. Numerical parameters of the main-text simulations in Figs. 1–3 and 5–7. The regime map uses the grid specified in Sec. SIV; the channel analysis uses the transverse packet width specified in Sec. SV.

| Parameter | Physical value |
|---|---|
| Fermi velocity $v_{\mathrm{F}}$ | $10^6$ m s$^{-1}$ |
| Domain $L_x = L_y$ | 3.29 $\mu$m |
| Grid $N_x = N_y$ | 1024 |
| Spacing $\Delta x = \Delta y$ | 3.21 nm |
| Time step $\Delta t$ | 0.263 fs |
| Maximum propagation time | 13.16 ps |
| Central energy $E_{\mathrm{F}}$ | 55 meV |
| Packet widths $\sigma_x = \sigma_y$ | 132 nm |
| Contact and wall strips $W_x, W_y$ | 165 nm |
| Mask strength $S$ | 0.08 |
| Wall mass $M_{\mathrm{wall}}$ | 500 meV |
| Residual-probability threshold | $10^{-3}$ |

## SIII. ROLE OF INJECTION COLLIMATION

The geometric mechanism acts on the transverse-momentum content of the injected state, so the strength of the directional asymmetry is set by how tightly the contact collimates the beam. We quantify this dependence here, both to characterize the effect and to connect the simulated injector to a realistic finite contact.

The angular acceptance of the injected Gaussian beam follows directly from the wave-packet defined in Sec. SI [Eq. (S1)]. With $\sigma_x = \sigma_y \equiv \sigma$ and the central momentum aligned along $\hat{\boldsymbol{x}}$ ($k_{x,0} = k_0$, $k_{y,0} = 0$), the transverse envelope at the injection plane reduces to

$$\psi(y) \;\propto\; \exp\left[-\frac{y^2}{4\sigma^2}\right], \tag{S7}$$

so that the transverse position probability density is $|\psi(y)|^2 \propto \exp[-y^2/(2\sigma^2)]$ and $\sigma$ is the strict one-$\sigma$ standard deviation of the transverse position. Fourier transforming Eq. (S7) yields a Gaussian in $k_y$,

$$\tilde{\psi}(k_y) \;\propto\; \exp[-\sigma^2 k_y^2] \;\;\Rightarrow\;\; |\tilde{\psi}(k_y)|^2 \;\propto\; \exp[-2\sigma^2 k_y^2], \tag{S8}$$

whose one-$\sigma$ width is

$$\sigma_{k_y} = \frac{1}{2\sigma}, \tag{S9}$$

saturating the Heisenberg minimum $\sigma\,\sigma_{k_y} = \frac{1}{2}$. The plane-wave components of the packet obey $k_y = k_0 \sin\theta$, so in the small-angle regime $\sigma_{k_y} \ll k_0$ relevant to this work, the incidence angle inherits a Gaussian distribution centered at $\theta = 0$ with one-$\sigma$ width

$$\sigma_\theta = \frac{\sigma_{k_y}}{k_0} = \frac{1}{2\,\sigma\,k_0}. \tag{S10}$$

We adopt this strict one-$\sigma$ definition, $\Delta\theta \equiv \sigma_\theta = 1/(2\sigma k_0)$, on the top axis of Fig. S1; the $2\sigma$ window (the characteristic full angular spread) is twice as wide.

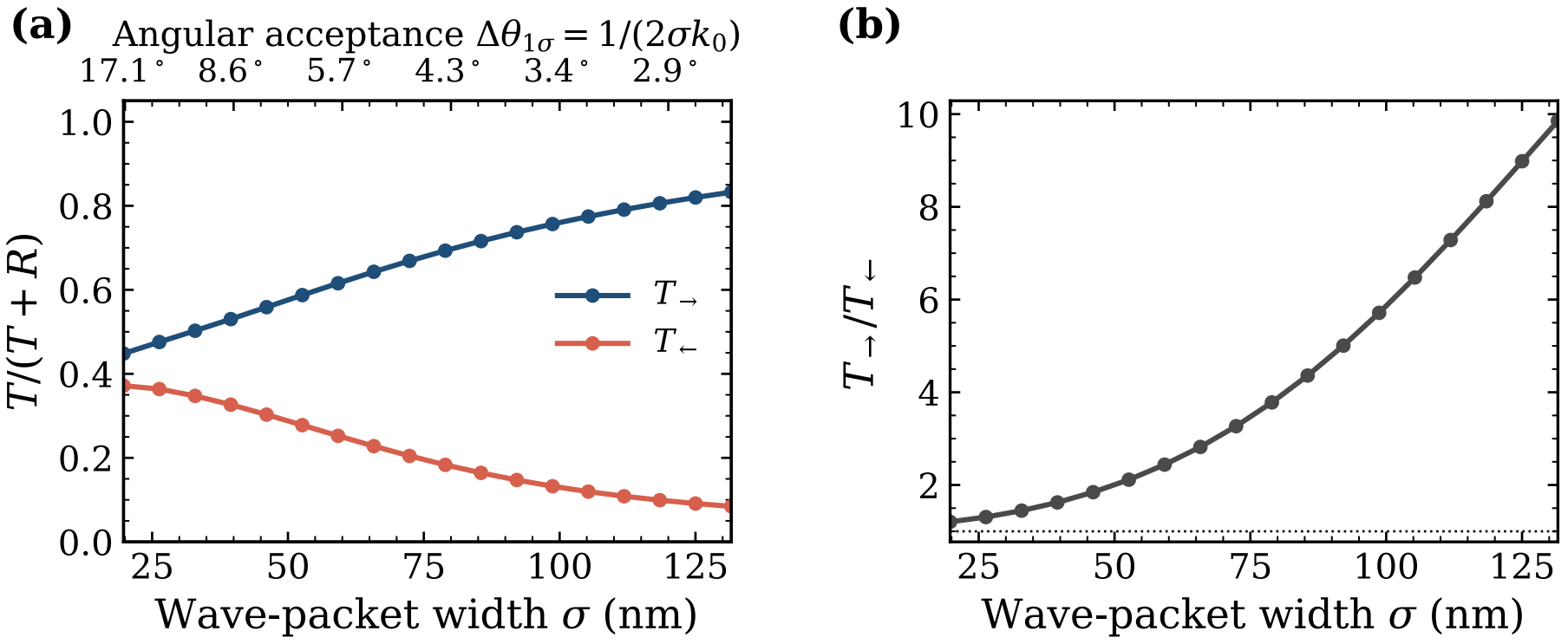

FIG. S1. **Wave-packet collimation enhances the coherent directional asymmetry. (a)** Forward ($T^{\rightarrow}$) and backward ($T^{\leftarrow}$) coherent transmission through the right-angle V–O barrier as a function of the Gaussian wave-packet width $\sigma$ (bottom axis) and the corresponding one-$\sigma$ angular acceptance $\Delta\theta$ (top axis). **(b)** Forward/backward transmission ratio $T^{\rightarrow}/T^{\leftarrow}$ over the same $\sigma$ range. Parameters: $\zeta_y = 0.35$, $E_F = 55\ meV$ and $V_0 = 50\ meV$, as in Figs. S2 and S3 and inside the directional window of Figs. 3 and 5 of the main text; geometry as in Fig. 1 of the main text; only $\sigma$ is varied. Data in both panels are for the $K$ valley.

As $\sigma$ increases at fixed Fermi energy and barrier height (Fig. S1), the injection becomes more collimated; the forward channel then saturates toward unity while the backward channel is increasingly suppressed, producing a monotonic rise of the forward/backward ratio, from $\approx 1.2$ at $\sigma \approx 20\ nm$ ($\Delta\theta \approx 17°$) to $\approx 10$ at $\sigma \approx 132\ nm$ ($\Delta\theta \approx 2.6°$). The $\sigma = 132\ nm$ width used in Figs. 1–3 and 5–7 of the main text corresponds to $\Delta\theta \approx 2.6°$ at $E_F = 55\ meV$, well inside the collimation-dominant regime.

The packet width describes the effective transverse extent of the injected beam after collimation. A broad Gaussian models a diffraction-limited coherent beam, while reducing $\sigma$ increases its angular spread according to Eq. (S10). Even the broadest Gaussian acceptance in Fig. S1, $\Delta\theta \approx 17°$, retains a measurable directional contrast. Absorptive pinhole collimators in ballistic graphene produce beams with an angular width of $\lesssim 18°$,[8] and imaging experiments resolve electron flow from collimating contacts.[9] The experimental beam width and angular distribution depend on the injector together with its collimator. Adiabatic widening can collimate carriers emitted by a quantum point contact.[10] The results in Fig. S1 quantify transmission for the specified Gaussian preparation; complete reciprocal sets of incoming modes with matched weights obey the channel-sum relation discussed in Sec. SV.

## SIV. ENERGY DEPENDENCE AND OPERATING REGIMES

For isotropic dispersion, the relative electron-optical index across a gate-defined interface is $n = |1 - V_0/E_F|$. Let the oblique face make an angle $\alpha$ with the vertical. A normally incident beam entering this face has incidence angle $\alpha$ relative to its normal. Conservation of tangential momentum requires $\sin\theta_{in} = \sin\alpha/n$ inside the barrier. Thus, the oblique-first direction is evanescent for $n < \sin\alpha$. In the vertical-first direction, normal entry preserves the propagation direction inside the barrier, and the oblique exit requires $\sin\theta_{out} = n\sin\alpha$. This direction is blocked for $n > 1/\sin\alpha$. These conditions describe the central ray; the finite angular and energy widths of the packet broaden the transitions.

For $\alpha = 45°$, the thresholds are $n = 1/\sqrt{2}$ and $n = \sqrt{2}$. Over the plotted interval $-1 \le V_0/E_F \le 2$, they give the three boundaries $1-\sqrt{2}$, $1 - 1/\sqrt{2}$ and $1 + 1/\sqrt{2}$. Oblique-first transmission dominates below $1-\sqrt{2}$, while vertical-first transmission dominates between $1 - 1/\sqrt{2}$ and $1 + 1/\sqrt{2}$. Both directions transmit in the intervening interval and above $1 +$ $1/\sqrt{2}$ within the plotted range. For the 8.6° facet of the main geometry at $E_{\mathrm{F}} = 55$ meV, the oblique-first blocking interval is $47$ meV $< V_0 <$ $63$ meV. More generally, its center is $V_0 = E_{\mathrm{F}}$ and its width is $2E_{\mathrm{F}}\sin\alpha$.

Figure 4 of the main text uses $E_{\mathrm{F}} = 65$ meV, isotropic dispersion and a 45° facet. The 234 values of $V_0$ cover $-65$ to $129.32$ meV, corresponding approximately to $-1 \leq V_0/E_{\mathrm{F}} \leq 2$. The domain is $L_x = 3.29$ $\mu$m, $L_y = 1.316$ $\mu$m on a $960 \times 384$ grid. The triangular barrier has vertices $(-461, -658)$ nm, $(-461, 658)$ nm and $(855, -658)$ nm, and an interface smoothing length of 13.2 nm. The time step is 0.263 fs and the maximum propagation time is 13.16 ps. The side walls are reflecting, with wall mass 500 meV. Figure 4(c–e) shows the probability density for $V_0/E_{\mathrm{F}} = -0.70, -0.10$ and $+0.55$, one gate value in each of the regimes of Fig. 4(a,b), at 1.05, 1.18 and 1.18 ps (V→O) and 1.58, 1.58 and 1.18 ps (O→V) after launch. The transmission values in the panels are those of Fig. 4(a).

The coherent angular source is formed by simultaneously injecting Gaussian spinor wave-packets with different central propagation angles. For each injection direction $d$, its initial state is

$$\Psi_d(\boldsymbol{r}, 0) = \mathcal{N}_d \sum_{m=1}^{11} \psi_{d,m}(\boldsymbol{r}),$$
$$\psi_{d,m}(\boldsymbol{r}) = g_d(\boldsymbol{r})\, e^{i\boldsymbol{k}_{d,m}\cdot\boldsymbol{r}} \chi_{+,K}(\boldsymbol{k}_{d,m}),$$

where $g_d$ is the Gaussian envelope of Eq. (S1), centered at $x_d = \mp 987$ nm and $y_d = 0$ for V→O and O→V injection, respectively. All eleven components have equal amplitudes and zero relative phases in this plane-wave convention, with $\sigma_x = \sigma_y = 132$ nm and central energy 65 meV. For isotropic dispersion, the central wave-vector magnitude is $k_{\mathrm{F}} = E_{\mathrm{F}}/(\hbar v_{\mathrm{F}}) = 0.0988\ \mathrm{nm}^{-1}$, and the propagation direction coincides with the wave-vector direction. The component wave vectors are

$$\boldsymbol{k}_{\mathrm{V}\to\mathrm{O},m} = k_{\mathrm{F}}(\cos\theta_m, \sin\theta_m), \qquad \boldsymbol{k}_{\mathrm{O}\to\mathrm{V},m} = k_{\mathrm{F}}(-\cos\theta_m, \sin\theta_m),$$

with $\theta_m = 0°, \pm 1.560°, \pm 4.679°, \pm 7.798°, \pm 10.917°$ and $\pm 14.036°$. The centers span 28.072°, and each component retains its Gaussian momentum width, corresponding to an angular standard deviation of approximately 2.20°. These parameters define the nominal 30° source range.

The normalization includes the overlap between components,

$$\mathcal{N}_d^{-2} = \sum_{m,n=1}^{11} \int d^2r\, \psi_{d,m}^{\dagger}(\boldsymbol{r}) \psi_{d,n}(\boldsymbol{r}).$$

The complete initial state is propagated as one two-component wavefunction under the Hamiltonian and evolution equation given above. Its density and contact-absorbed probability therefore retain the interference terms between angular components. The packets form a coherent single-particle state with a finite angular distribution.

The curves report cumulative contact transmission over propagation times up to 13.16 ps. The shaded bands in Fig. 4(a) show the uncollected probability $P_{\mathrm{rem},d} = 1 - T_d - R_d$ for direction $d$, drawn on top of the collected transmission $T_d$. This probability resonates in the channel through multiple reflections between the side walls and the barrier. Since it may still leave through either contact, the band height is the largest possible increase of $T_d$ at longer collection times. The largest values of $P_{\mathrm{rem},d}$ over the $V_0$ range are 3.63% for O→V and 5.58% for V→O. The corresponding bounds on the ratio in Fig. 4(b) are

$$\frac{T_{O\to V}}{T_{V\to O} + P_{\mathrm{rem},V\to O}} \leq \frac{T_{O\to V}^{\mathrm{final}}}{T_{V\to O}^{\mathrm{final}}} \leq \frac{T_{O\to V} + P_{\mathrm{rem},O\to V}}{T_{V\to O}}.$$

The uncollected probability therefore leaves the strong directional contrast and its gate-controlled reversal intact.

## SV. CHANNEL-RESOLVED TRANSMISSION

For flux-normalized channels of the Hamiltonian in Eq. (1) of the main text, time reversal $\Theta = \sigma_z \mathcal{K}$ exchanges the valleys and reverses the momentum. The reciprocal channel relation is

$$|t_{R\to L}^{K'}(-k_y' \to -k_y)|^2 = |t_{L\to R}^{K}(k_y \to k_y')|^2. \qquad \text{(S11)}$$

For one fixed incident channel $i$, the directional totals are

$$T_i^F = \sum_j |t_{ji}|^2, \qquad T_i^B = \sum_j |t_{ij}|^2. \tag{S12}$$

These column and row sums can differ. The complete sums over incoming modes are equal by unitarity for a lossless two-terminal scattering problem. Collecting every outgoing angle preserves this distinction because it completes only the outgoing-channel sum. The complete incoming-channel sum is treated analytically here.

The channel analysis uses the right-angle barrier of Fig. 1 in the main text, $V_0 = 50$ meV and $E_F = 55$ meV. We take $\sigma_x = 132$ nm and $\sigma_y = 460$ nm to sharpen the transverse-momentum resolution. The transmitted field is sampled on the drain side over $526$ nm $< |x| < 1.448\ \mu$m. Its Fourier components are weighted by the outgoing longitudinal speed $|v_x| = v_F|k_x|/|\boldsymbol{k}|$ and accumulated over time. In discrete form, the probability assigned to each transverse-momentum bin is

$$P(k_y) = \sum_t \frac{\Delta t_{\text{save}}}{W} \sum_{k_x \text{ outgoing}} |v_x| \mathcal{N}_{\text{FFT}} \sum_{a=1}^{2} |\tilde{\psi}_a(k_x, k_y, t)|^2, \tag{S13}$$

where $W$ is the sampled longitudinal width and $\mathcal{N}_{\text{FFT}}$ is the Parseval normalization. The spectrum estimate is $T_{\text{spectrum}} = \sum_{k_y} P(k_y)$. The bin spacing is $2\pi/L_y = 0.00191\ \text{nm}^{-1}$. Lobe weights are summed over $|k_y - k_c| < 0.00608\ \text{nm}^{-1}$, where $k_c$ is the peak-bin position. For the tilted forward beam, $k_c = 0.01146\ \text{nm}^{-1}$ and the lobe centroid is $0.01085\ \text{nm}^{-1}$; the reversed-channel output is centered near zero.

For the tilted cone, normal $K$ injection from the left produces a deflected output at $k_y \approx 0.0108\ \text{nm}^{-1}$. Its reciprocal preparation is $K'$ injection from the right along the reversed peak-bin direction, $k_y = -0.0109\ \text{nm}^{-1}$, at the same central energy. The momentum magnitude follows from the electron-band eigenenergy,

$$|\boldsymbol{k}_0| = \frac{E_F}{\hbar[v_F + \tau w_y \sin\phi_0]}, \tag{S14}$$

and is $0.0798\ \text{nm}^{-1}$ for this reverse preparation. Normal incidence has $k_y = 0$, for which the transverse tilt leaves $k_0 = E_F/(\hbar v_F)$. Reversing the transmitted channel therefore requires both its reversed direction and the momentum magnitude appropriate to that direction at the same energy.

Table S2 separates contact absorption, the flux-spectrum estimate and the selected lobe weight. For the tilted cone, the forward and reversed-channel contact transmissions are $0.974$ and $0.971$, while the two lobe weights are $0.89$ and $0.90$. These integrated finite-packet weights agree at the percent level, consistent with reciprocal channel conversion. They describe channel groups with finite angular and energy widths. The spectrum estimator differs from contact absorption by a few $10^{-3}$. For the $\sigma_x = \sigma_y = 132$ nm packet of the main-text calculations, the normally incident $K$-forward and $K'$-backward contact transmissions are 0.831 and 0.311. The channel analysis thus identifies transmitting reverse channels even when the normally incident reverse preparation is suppressed.

TABLE S2. Transmission at $V_0 = 50$ meV and $E_F = 55$ meV for the channel preparations in Fig. S2. The lobe windows are defined in Sec. SV.

| Dispersion | Preparation | $T_{\text{contact}}$ | $T_{\text{spectrum}}$ | Lobe weight |
|---|---|---|---|---|
| Untilted | Forward, normal ($K$) | 0.974 | 0.970 | 0.967 |
| | Backward, normal ($K$) | 0.084 | 0.083 | 0.082 |
| | Backward, reversed channel ($K$) | 0.967 | 0.963 | 0.963 |
| Tilted | Forward, normal ($K$) | 0.974 | 0.970 | 0.893 |
| | Backward, normal ($K'$) | 0.257 | 0.256 | 0.255 |
| | Backward, reversed channel ($K'$) | 0.971 | 0.967 | 0.903 |

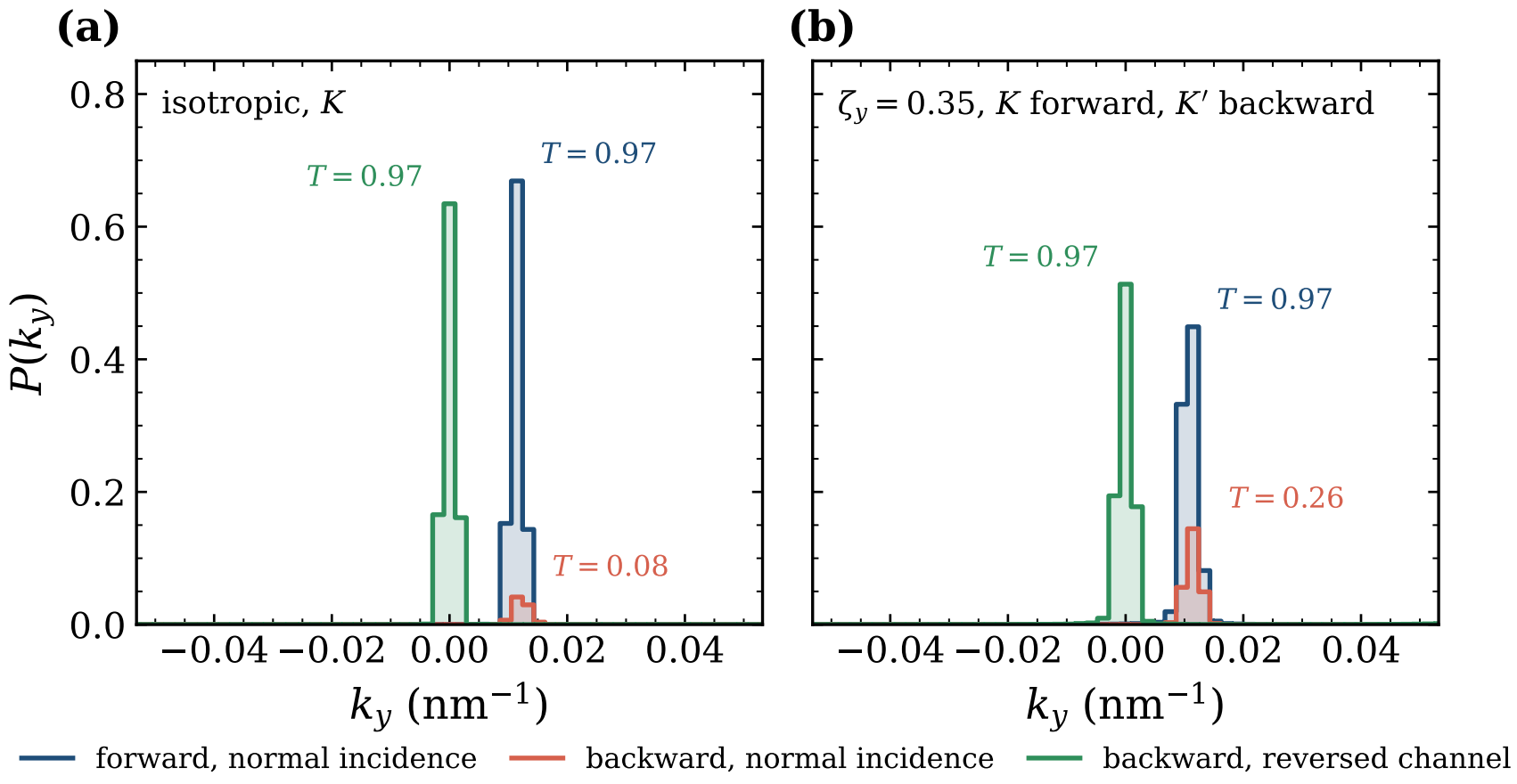

FIG. S2. **Channel-resolved transmission.** Flux-weighted transverse-momentum spectra of the transmitted field for the normally incident forward beam (blue), normally incident backward beam (orange), and backward beam injected along the reversed deflected channel at the same central energy (green). Left: isotropic dispersion, valley $K$. Right: $\zeta_y = 0.35$, $K$ forward and $K'$ backward, with $|k| = 0.0798\ \mathrm{nm}^{-1}$ for the reversed-channel preparation. The forward output is deflected, while the reversed-channel output concentrates near $k_y = 0$. The range $|k_y| < 0.0532\ \mathrm{nm}^{-1}$ displays the central lobes; 5–6% of the tilted spectra lie outside this range.

## SVI. BOUNDARY-CONDITION AND DETECTOR CONTROLS

The absorbing contacts span the full channel width, and the transverse walls are lossless. Transmission and reflection therefore include the full outgoing angular distribution. The barrier-free reference and the rectangular and isosceles controls use the same absorbers as the right-angle barrier. The forward and backward normalized transmissions agree within 0.65% for the rectangular control and 0.16% for the isosceles control in each valley over the applied potential range. For the parallelogram, the opposite-valley pairs, $T_K^{\rightarrow} = T_{K'}^{\leftarrow}$ and $T_{K'}^{\rightarrow} = T_K^{\leftarrow}$, agree within 0.7%, and the valley-averaged transmissions $\bar{T} = (T_K + T_{K'})/2$ of the two injection directions agree within 0.11%, as reported in Sec. II E of the main text. For two compared normalized transmissions $a$ and $b$, the relative difference is $100|a - b|/\min(a, b)$ in percent.

### Injection position

We displace the normally incident packet across the same right-angle barrier at $E_\mathrm{F} = 55$ meV and $V_0 = 50$ meV, with isotropic dispersion and $\sigma_x = \sigma_y = 132$ nm. Table S3 and Fig. S3 show the contact-collected $K$-valley transmissions. The forward transmission remains high at all injection positions, while the backward transmission increases toward the thin end of the wedge. The directional ratio ranges from 2.1 to 7.6.

In the tilted case, the transverse tilt produces a drift of approximately 244 nm before the first encounter with the barrier, corresponding to a local-width change of approximately 75 nm between the two displaced trajectories. The packet itself samples local widths of 210–290 nm across its $\pm 2\sigma$ extent. The position scan shows that the interface-order effect persists across this variation, while the local barrier width modulates the collected transmission.

TABLE S3. Injection-position dependence for the $K$ valley. $y_0$ is the initial transverse displacement in nm.

| $y_0$ (nm) | $T^{\rightarrow}$ | $T^{\leftarrow}$ | $T^{\rightarrow}/T^{\leftarrow}$ |
|---|---|---|---|
| −987 | 0.804 | 0.106 | 7.6 |
| −658 | 0.798 | 0.108 | 7.4 |
| −329 | 0.803 | 0.113 | 7.1 |
| 0 | 0.813 | 0.127 | 6.4 |
| 329 | 0.831 | 0.162 | 5.1 |
| 658 | 0.865 | 0.248 | 3.5 |
| 987 | 0.914 | 0.442 | 2.1 |

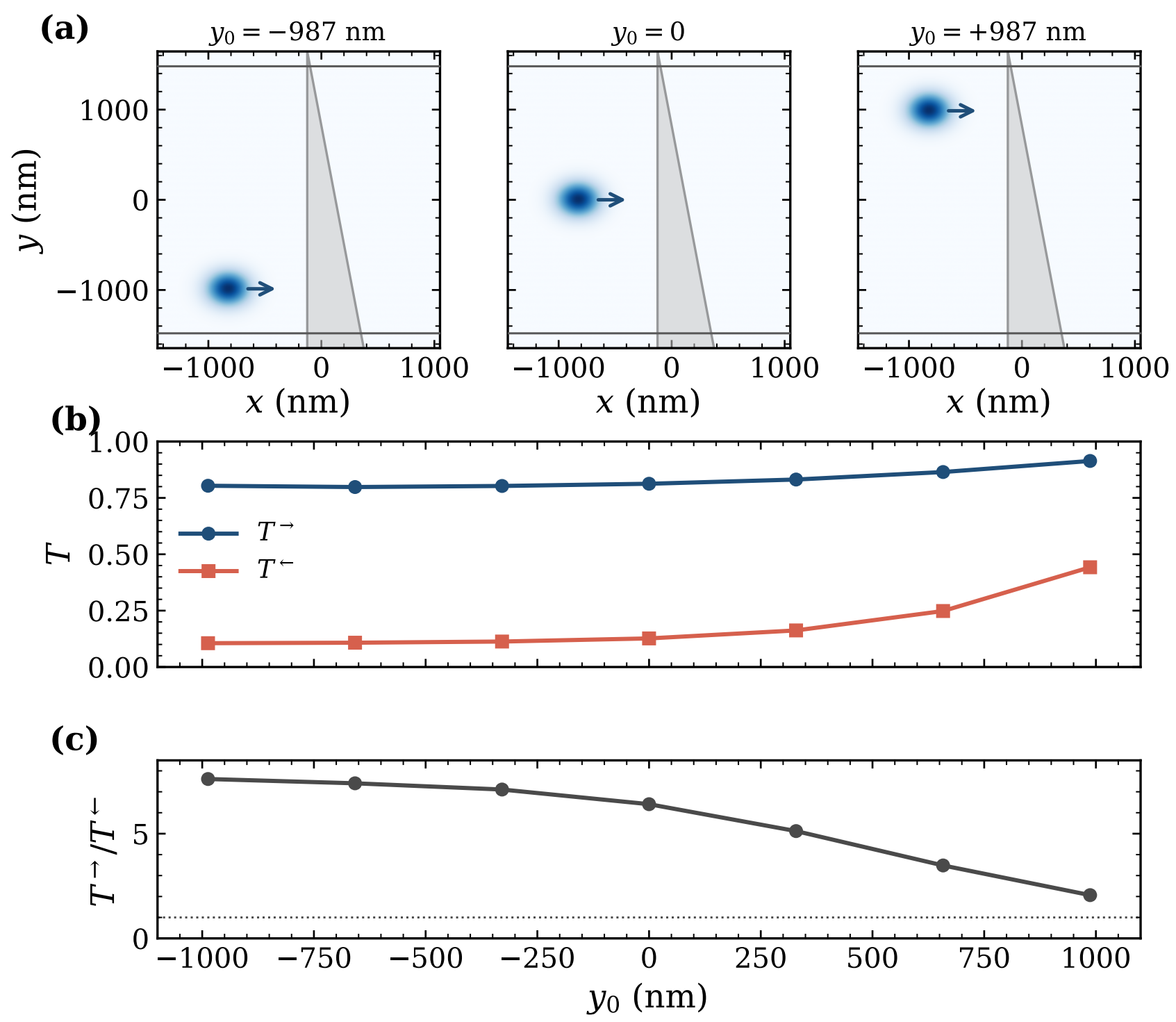

FIG. S3. **Dependence on injection position.** Contact-collected forward and backward $K$-valley transmission and their ratio as the initial beam center is displaced across the right-angle barrier. Parameters: $E_{\mathrm{F}} = 55$ meV, $V_0 = 50$ meV, isotropic dispersion and $\sigma_x = \sigma_y = 132$ nm. The directional contrast persists over the scanned range.

## Side-wall collection

For a normally incident Gaussian packet at the 45° barrier of Fig. 4 in the main text, with $V_0/E_{\mathrm{F}} = -0.5$, replacing the reflecting side walls with absorbing probe contacts gives a four-terminal configuration that removes deflected flux. The O→V and V→O transmissions change from 0.984 and 0.021 with reflecting walls to 0.445 and $7 \times 10^{-4}$ with absorbing side contacts. The directional transport persists and becomes stronger when the deflected flux is removed. The return of flux from reflecting walls therefore changes the collected probabilities while preserving the directional asymmetry in this control.